\documentclass[11pt]{article}
\usepackage{amssymb}
\usepackage{epsfig}
\begin{document}
\begin{titlepage}
\title{Detecting planets around stars in nearby galaxies}

\author{{G. Covone$^{1}$, R. de Ritis$^{1,2}$, M. Dominik$^{3}$, A. A. Marino$^{2,4}$} \\
{\em \small $^1$Dipartimento di Scienze Fisiche, Universit\`{a} di 
Napoli,} \\
{\em \small Mostra d'Oltremare pad. 20-80125 Napoli, Italy;} \\
{\em \small $^2$Istituto Nazionale di Fisica Nucleare, Sezione di Napoli,} \\
{\em \small Complesso Universitario di Monte S. Angelo, Via Cinzia, Edificio G 80126 Napoli, Italy;} \\
{\em \small $^3$ Kapteyn Astronomical Institute, Postbus 800, NL-9700 AV Groningen, The Netherlands;} \\
{\em \small $^4$ Osservatorio Astronomico di Capodimonte, Via Moiariello, 16, I-80131 Napoli, Italy.}}
\date{}             
\maketitle    

\begin{abstract}

The only way to detect planets around stars at distances $\gtrsim$
several kpc is
by (photometric or astrometric) microlensing ($\mu$L) observations.
In this paper, we show that the capability of 
photometric $\mu$L{} extends to the detection of signals caused by
planets around stars
in nearby galaxies (e.g.\ M31) and that there is no other method 
that can achieve this. 
Due to the large crowding, $\mu$L{} experiments towards M31 can only
observe the high-magnification part of a lensing light curve.
Therefore, the dominating channel for $\mu$L{} signals by planets is
in distortions near the peak of high-magnification events
as discussed by Griest \& Safizadeh (\cite{GS98}).  
We calculate the probability to detect planetary anomalies 
for $\mu$L{} experiments towards M31
and find that jupiter-like planets around stars in M31 can be detected.
Though the 
characterization of the planet(s) involved in this signal will be difficult,
the absence of such signals can yield strong constraints on the abundance
of jupiter-like planets.

\end{abstract}\vspace{20. mm}

\noindent e-mail addresses:
covone@na.infn.it \\
deritis@na.infn.it \\
dominik@astro.rug.nl \\
marino@na.astro.it \\
	      \vfill
	      \end{titlepage}

\section{Introduction} 

The existence of `other worlds' has always been one of the most
discussed topics in the history of philosophy and science. The
question has fascinated researchers since more than 2000 years, but the
first attempt in modern astronomy to discover extrasolar planets was
given by Huyghens (\cite{Huyghens}), in the XVII century.
One had to wait nearly another 300~years until the first extrasolar planets
have been discovered (Mayor \& Queloz~\cite{planet1}; Marcy \& Butler~\cite{planet2}),
namely by observing the radial velocity of the parent star by Doppler-shift
measurements. All of the confirmed detections of extrasolar planets so far
result from this technique and $\sim 20$ planets have been found
(Schneider~\cite{planetencyc}).

Already in 1991, Mao
\& Paczy\'nski (\cite{MP}) have pointed out that not only a (dark) foreground star
that passes close to the line-of-sight of an observed luminous background source star
yields a detectable variation in the observed light of the source star but also
a planet around the foreground (lens) star can significantly modify the observed
light curve.
Gould \& Loeb (\cite{Gould92}) have shown that there is a significant probability to
detect jupiter-mass and saturn-mass planets around stars in the Galactic disk
that act as microlenses by magnifying the light of observed stars in the Galactic bulge.
Bennett \& Rhie (\cite{Bennett96}) have pointed out that the capability of 
detecting planets by this photometric microlensing ($\mu$L{}) technique
extends to earth-mass planets, where the limit is given by the finite size of the source
stars.

Contrary to all techniques employed or suggested 
to search for planets, photometric $\mu$L{} does not favour nearby objects.
This makes it the unique technique to search for planets around stars at distances 
larger than a few kpc. Moreover, for disk lenses and bulge sources, a separation between planet
and parent star of 2--6~AU is favoured, making it an ideal method to look for jupiter-like systems.
Since the parent star
of the planet acts as a gravitational lens only through its gravitational field,
there is no luminosity bias for the parent stars that are generally not even seen.
Moreover, it is the only method to
discover Earth-like planets from ground-based 
observations.\footnote{In 1992, Earth mass
objects have been discovered around the pulsar PSR1257+12 (Wolszczan \& Frail~\cite{Wols1};
Wolszcan~\cite{Wols2})
through time-delay measurements. The discovery is undoubtful, but the
very nature of these objects is completely unknown: it is difficult,
at the moment, to conciliate this discovery with our picture of
planetary systems. A precise definition of a planet is a subtle
question (see Marcy \& Butler~\cite{Marcy98}).}  

Several teams have started to look for planetary anomalies in
$\mu$L{} light curves with monitoring programs that perform frequent and 
precise observations, namely
PLANET (Albrow et al.~\cite{PL}; Dominik et al.~\cite{PL2}), MPS (Rhie et al.~\cite{MPS}), and
MOA (Hearnshaw et al.~\cite{MOA}). All these teams rely on the microlensing 'alerts'
issued by teams that undertake surveys of $\sim 10^{7}$ stars:
OGLE (Udalski et al.~\cite{OGLE}), 
MACHO\footnote{MACHO will discontinue its operation by the end of 1999.} 
(Alcock et al.~\cite{MACHO1,MACHO2}), and EROS (Palanque-Delabrouille et al.~\cite{EROS}).

While most of these alerts are on Galactic bulge stars,
MACHO and EROS also observe(d) fields towards the
Magellanic Clouds. However, the number of
events towards SMC and LMC comprises  
only 5--10\% of the total number of events. In addition to detecting planets
around stars in the Galactic disk (typically at $4~\mbox{kpc}$ distance) one
could also think of detecting planets around stars in the Magellanic Clouds
(at $\sim 50~\mbox{kpc}$ distance). However, in addition to the relative small number
of detected events, finite source effects play a much more prominent role for
lensing of stars in the Magellanic Clouds by stars in the Magellanic Clouds
than for lensing of Galactic bulge stars by Galactic disk stars 
(Sahu~\cite{SahuNat}) resulting in a dramatic decrease in the probability to
detect planetary signals.

Safizadeh et al. (\cite{Safplan}) have pointed out that planets around disk stars can
also be detected by looking at the shift of the light centroid of observed source stars caused
by microlensing of disk stars and surrounding planets with upcoming space interferometers that allow
to measure astrometric shifts at the $\mu$as level. Contrary to photometric $\mu$L{}, the observed
signal of this
'astrometric $\mu$L{}' technique decreases with the distance of the lenses 
(e.g.\ Dominik \& Sahu~\cite{Sahu}). With $\mu$as-astrometry, jupiter-mass planets can only be 
detected for distances up to $\lesssim 30~\mbox{kpc}$. 
This leaves photometric $\mu$L{} as the only method ever capable of detecting 
planets in nearby galaxies like M31. 

In contrast to microlensing observations towards the Galactic bulge and the Magellanic Clouds,
a large number of source stars fall onto the same pixel of the detector for observations towards
M31. However, it is still possible to detect $\mu$L{} events even in unresolved star fields 
(Baillon et al.~\cite{Baillon}; Gould~\cite{Gould96}). 
Since standard photometric methods cannot be used to reveal $\mu$L{} events,
new techniques
have been developed: super-pixel photometry (Ansari et al.~\cite{Ansari97})
and difference image photometry (Tomaney \& Crotts~\cite{TC}; Alard \& Lupton~\cite{Alard98}).
These techniques are used for the $\mu$L{} searches towards M31 
as carried out by
the Columbia-VATT search (Crotts \& Tomaney~\cite{CT}), AGAPE
(Ansari et al.~\cite{Ansari97}), SLOTT-AGAPE (Bozza et al.~\cite{SLOTT}), and MEGA
(Crotts et al.~\cite{MEGA}).

In this paper we investigate the possibility 
to detect planets around stars in M31 with experiments that make use of either of these techniques.
By searching for
planets (or, at least, brown dwarfs) even in other galaxies, the limit for planet detection
is further pushed towards larger distances.

The paper is organized in the following way: in Sect.~2, we discuss the characteristics
of microlensing signals caused by planets.
In Sect.~3, the
conditions for detecting anomalies in light curves of M31 
are discussed. 
In Sect. 4, we calculate the probability to detect planetary signals in M31,
and in Sect.~5, we discuss the extraction of planetary parameters.
Finally, in Sect.~6, we summarize and conclude.

\section{Microlensing signals of planets}
\label{miclens}

A microlensing event occurs if a massive lens object with mass $M$ located at
a distance $D_{\rm L}$ from the observer passes close to the line-of-sight
towards a luminous source star at the distance $D_{\rm S}$ from the observer.
Let $u$ denote the angular separation betwen lens and source in units
of the angular Einstein radius
\begin{equation}
\theta_{\rm E} = \sqrt{\frac{4GM}{c^2}\,\frac{D_{\rm S} - D_{\rm L}}{D_{\rm L}\,D_{\rm S}}}\,.
\end{equation}
For the 'standard model' of $\mu$L{}, i.e. point-like sources and lenses,  
the magnification $\mu$ is then given by (Paczy{\'n}ski~\cite{Pac86})
\begin{equation}
\mu(u) = \frac{u^2+2}{u\,\sqrt{u^2+4}}\,.
\end{equation}
If one assumes uniform rectilinear motion between lens and source with the
relative proper motion $\mu$, one has
\begin{equation}
u(t) = \sqrt{u_0^2+\left(\frac{t-t_0}{t_{\rm E}}\right)^2}\,,
\end{equation}
where $t_{\rm E} = \theta_{\rm E}/\mu$, $u_0$ gives the impact parameter, and $t_0$ gives
the time of the smallest separation between lens and source.
This means that one observes a light curve $\mu(u(t))$ that has the form 
derived by
Paczy{\'n}ski (\cite{Pac86}), the so-called Paczy{\'n}ski curve.
For recent and complete reviews of the theory of microlensing 
and of the observational results we further refer the to the works of
Paczy{\'n}ski (\cite{Pac-rev}),
Roulet \& Mollerach (\cite{Roulet97}), and
Jetzer (\cite{Jetzer98}). 

More sophisticated models of the lens and the source include the finite
source and the binarity (or multiplicity) of these objects.
For such models, 
the light curves can differ significantly from 
Paczy{\'n}ski curves.

If one neglects the binary motion,
a binary lens is characterized by two
parameters, the mass ratio between the lens objects $q$ and their
instantanous angular separation $d$, measured in units of $\theta_{\rm E}$.
The model of a binary lens includes the configuration of a star that is
surrounded by a planet. In the following, we let $M$ denote
the mass of the more massive object (star), 
while $m$ denotes the mass of the less massive object (planet) and
$q = m/M < 1$. This means that $\theta_{\rm E}$ refers to the mass
$M$ of the more massive object. 

For any mass ratio $q$, the caustics of a binary lens can
show three different topologies (Schneider \& Wei{\ss}~\cite{SchneiWei};
Erdl \& Schneider~\cite{Erdl}) depending on the separation $d$: 
For 'wide binaries' there are two disjoint diamond-shaped caustic near the 
positions of each of the lens objects,
for 'intermediate binaries' there is only one caustic with 6 cusps,
and for 'close binaries' there is one diamond-shaped caustic near the
center-of-mass and two small triangular shaped caustics.
As $q \to 0$, the region of intermediate binaries vanishes as $q^{1/3}$
and the transition close-intermediate-wide occurs at $d=1$ 
(Dominik~\cite{Dominik99}).
This means that for planets,  
one has a 'central caustic' near the star and either a diamond-shaped caustic
(for $d>1$) or two triangular shaped caustics (for $d < 1$) at the position
that had an image under the lens action of the star, considered at the position
of the planet.
 We will refer to the latter caustic(s) as 'planetary
caustic(s)'.

Since the caustics are small and well-separated, the light curve mainly
follows a Paczy{\'n}ski curve and is only locally distorted by either of
the caustics.
This allows us to distinguish two main types of anomalies in the light curve,
namely the events affected by the central caustic (type I), and the ones 
affected by one of the planetary caustics (type II).

To produce a Type I anomaly, the source has to pass the lens star with
a small impact parameter, say $u_0 \lesssim 0.1$. Unless the source size is
larger than variations in the magnification pattern, type I anomalies occur
in high-magnification events ($\mu \simeq 1/u$ for $u \ll 1$).  
Moreover, the anomaly occurs near the maximum of the underlying
Paczy{\'n}ski curve. 
Griest \& Safizadeh (\cite{GS98}) have pointed out that for high-magnification
events, the probability to detect a planetary signal, namely as
type I anomaly, is very large.
In order to produce a high detection probability, the
central caustic is often elongated along the lens axis, so that
the magnification pattern is highly asymmetric around the lens star.
If there are $N$ planets with masses $m_i$ around the parent star 
with mass $M$,
they all perturbate the central caustic (Gaudi et al.~\cite{Naber98}),
where the effect is proportional to the mass ratios $q_i = m_i/M$
(Dominik~\cite{Dominik99}). Though in principle, one can obtain
information about the whole planetary system, the extraction of this 
information is non-trivial and the results are likely to be
ambiguous (Dominik \& Covone, in preparation).

Type II anomalies are produced when
the source passes close enough to the lens to produce a detectable
Paczy{\'n}ski curve ($u_0 \lesssim 1$), but not close enough to feel the effects of
the central caustic ($u_0 \gtrsim 0.1$), and also gets affected
by the planetary caustics, so that the source light beam will
also be deflected by the planet, and a perturbation of the Paczy\'nski curve
is produced at a time that depends 
on the angular separation between star and planet.
From this time and from the duration of the perturbations, mass
ratio $q$ and separation $d$ can be determined from high-quality observations,
unless the duration is strongly
influenced by the source size (Gaudi \& Gould~\cite{Gaudi97};
Dominik \& Covone, in preparation).

Experiments towards unresolved star fields in nearby galaxies
set very limiting conditions on the 
detection of $\mu$L{} events in general and on the detection
of anomalies in particular.
First, only the parts of the light curve that correspond
to large magnifications can be observed.
Second, anomalies can only be seen when they 
constitute very large deviations of the
received flux.
Therefore, all observed events are high-magnification events which
gives a lot of candidates to look for type I anomalies.
On the other hand, the background Paczy{\'n}ski curve for type II 
anomalies is not observed, and the planetary caustic has to be approached
very closely to produce a high magnification. Therefore, type II
anomalies are not likely to be detected in M31 experiments.

Griest \& Safizadeh (\cite{GS98}) have studied the influence of the finite
source size for type I anomalies. 
For sources in the Galactic bulge and lenses in the Galactic disk, they find
that the finite source size can be neglected even for giant sources
($R \sim 10~R_{\odot}$) for a parent star of solar-mass and a mass ratio
$q > 10^{-3}$. The characteristic quantity for the effect of the finite
source size is the ratio between source size and the physical size of the
angular Einstein radius at the position of the source 
\begin{equation}
r_E' = D_{\rm S}\,\theta_{\rm E} = \sqrt{\frac{4GM}{c^2}\,\frac{D_{\rm S}\,(D_{\rm S} - D_{\rm L})}
{D_{\rm L}}}\,.
\end{equation}
For lensing of bulge stars by disk stars, $D_{\rm S} \sim 8~\mbox{kpc}$ and
$D_{\rm L} \sim D_{\rm S}/2$, while for M31 sources and lenses, 
$D_{\rm S} \sim D_{\rm L} \sim 600~\mbox{kpc}$
and $D_{\rm S} - D_{\rm L} \sim 10~\mbox{kpc}$. Therefore $r_{\rm E}'$ is approximately the
same in the two cases and the estimates for the effect of the finite source size made
for bulge stars and disk lenses are also valid for M31 sources and lenses.

If the finite source size becomes non-negligible, the planetary signal is
suppressed. We therefore
restrict our discussion to planets with
mass ratio $q > 10^{-3}$, i.e.\
Jupiter-like planets around stars of solar-mass and systems with
larger mass ratio.

\section{Detectability of anomalies in M31 experiments}

For $\mu$L{} searches towards M31, each pixel of the detector
contains light from many 
unresolved stars.
There are several differences between classical microlensing
surveys (i.e. surveys on resolved stars)
and surveys towards unresolved star fields.

The first one concerns the photometric errors.
While in the classical regime, the photon noise is generally dominated
by the light from the lensed star, 
it is dominated by the flux from stars that are not lensed for observations
towards unresolved star fields. 
This means that the noise does not depend
on the magnification. 
A second important difference is that it is
impossible to determine the baseline flux of the lensed star. This means
that the actual magnification and
the Einstein time $t_{\rm E}$ of the event 
are not known.

Moreover, in surveys towards unresolved star fields, 
there is a natural selection bias
for the events with respect to the impact parameters and the luminosity of the
lensed sources (e.g.\ Kaplan~\cite{Kaplan97}): events that involve
lensing of giant stars and events with small impact parameters 
are preferred.

Searches of $\mu$L{} events towards unresolved star fields
(Crotts~\cite{Columbia};  
Baillon et al.~\cite{Baillon}), M31 in particular,  
have motivated the development of new photometric methods.
While the AGAPE team has implemented a 'super-pixel photometry' method
(Ansari et al.~\cite{Ansari97}; Kaplan 1998),
the Columbia-VATT team has used a 'difference image photometry' method
(Crotts \& Tomaney~\cite{CT}; Tomaney \& Crotts~\cite{TC}).  
Recently, Alard \& Lupton (\cite{Alard98}) have improved the latter 
method yielding the 'Optimal image subtraction' (OIS) technique.

The Columbia-VATT collaboration 
has found six candidate events towards M31 (Crotts \& Tomaney~\cite{CT}). 

AGAPE has observed 7 fields towards M31 in autumns
1994 and 1995, using the 2 meters telescope Bernard Lyot at the Pic du
Midi Observatory. Their data analysis has selected 19 microlensing
candidate events that are broadly consistent with Paczy\'nski curves.
Only two of them can be
retained as convincing candidates at the moment (Melchior~\cite{Mel}). 
One of these events shows a small but statistically significant deviation
from a Paczy{\'n}ski curve 
(Ansari et al.~\cite{Z1}).
This event
could be due to lensing of a binary source, or even to a binary lens.
There are too few data points to resolve the question, and other observations
are needed to confirm that the event is due to $\mu$L{} and not due to  
stellar variability. In any case, the
possibility to detect binary lens events towards unresolved star
fields has been demonstrated. 

This gives us some confidence that future $\mu$L{} searches
towards nearby galaxies
could not only detect binary-lens events, but also 
reveal Jupiter-like planets. 
From a general point of view, we expect 
a larger fraction of anomalous
microlensing events, since smaller
impact parameters are favoured  so that source trajectories are more likely
to pass through the more
asymmetric parts of the magnification pattern. However, the less
accurate photometry sets a severe limit on the detection of anomalies.
In the following, we determine how large an 
anomaly has to be in order to be detected in an M31 $\mu$L{} experiment.

The light in an observed pixel is composed of contributions from the
lensed star and many other unresolved stars. Since the light from the
lensed star is in general spread over several pixels, only a fraction $f$
of it is received on a given pixel. If $\mu$ denotes the magnification
of the lensed star, and $F_{\rm star}^{(0)}$ denotes its unlensed flux,
the flux variation on the pixel is given by
\begin{equation}
\Delta F_{\rm pixel} = (\mu - 1) f F_{\rm star}^{(0)},
\label{pixel}
\end{equation}
where $\mu$, $f$ and $F_{\rm star}^{(0)}$ are not observed individually.

Let us now consider an anomaly in an event, i.e.\ a deviation from a 
Paczy{\'n}ski curve. Let $\mu$ denote the magnification for the 
Paczy{\'n}ski curve and $\mu'$ the magnification for the anomalous
curve. The difference in the pixel flux variations is then given by
\begin{equation}
\Delta(\Delta F_{\rm pixel}) = (\mu' - \mu) f F_{\rm star}^{(0)}\,.
\end{equation}
This difference is detectable when it exceeds the rms fluctuation $\sigma_{\rm pixel}$ by a
factor Q, i.e.\
\begin{equation}
\mu' -\mu \geq Q \frac{\sigma_{\rm pixel}}{f F_{\rm star}^{(0)}}\,.
\end{equation}

One sees 
that the brighter the star the less the
magnification variation has to be in order to be detected. Thus, giant
stars are preferred as sources.

For $\mu \gg 1$, one obtains a detection threshold $\delta_{\rm th}$ for
anomalies with Eq.~(\ref{pixel}) as
\begin{equation}
\delta_{\rm th} \equiv 
\left|
\frac{\mu' - \mu}{\mu} \right|
_{\rm th} = Q \frac{\sigma_{\rm pixel}}
{\Delta F_{\rm pixel}}\,.
\label{mia2}
\end{equation}

\begin{table}
{\centering \begin{tabular}{|c|c|c|c|}
\hline 
\#&
$\sigma_{\rm pixel}$&
$(\Delta F_{\rm pixel})_{\rm max}$&
$\sigma_{\rm pixel}/(\Delta F_{\rm pixel})_{\rm max}$\\
\hline 
\hline 
1&
90&
850&
0.106\\
\hline 
2&
82&
680&
0.121\\
\hline 
3&
78&
1286&
0.061\\
\hline 
4&
99&
870&
0.114\\
\hline 
5&
85&
900&
0.094\\
\hline 
6&
44&
870&
0.050\\
\hline 
7&
46&
1200&
0.038\\
\hline 
8&
37&
1110&
0.033\\
\hline 
9&
53.5&
830&
0.064\\
\hline 
10&
73&
940&
0.077\\
\hline 
11&
100&
620&
0.094\\
\hline 
12&
56&
645&
0.121\\
\hline 
13&
101&
945&
0.107\\
\hline 
14&
63&
1320&
0.048\\
\hline 
15&
48&
790&
0.061\\
\hline 
16&
53&
600&
0.089\\
\hline 
17&
55&
780&
0.071\\
\hline 
18&
60&
807&
0.074\\
\hline 
19&
54&
860&
0.063\\
\hline 
\end{tabular}\par}
\caption{The rms fluctuation $\sigma_{\rm pixel}$ and the maximum flux variation
$(\Delta F_{\rm pixel})_{\rm max}$ for the 19 AGAPE candidate events towards M31,
analyzed using the super-pixel photometry method (Ansari et al.~\cite{Ansari97}).}
\end{table}

To obtain an estimate, we have a look at the values of $\sigma_{\rm pixel}$ and 
$(\Delta F_{\rm pixel})_{\rm max}$, i.e.\ $\Delta F_{\rm pixel}$ at the maximum,
for the 19 candidate events detected by AGAPE and analyzed using the
super-pixel photometry technique (Ansari et al.~\cite{Ansari97}).
This analysis has been made on $7
\times 7$ pixels squares, the so-called ``super-pixel'', which
correspond more or less to the average PSF dimension.
It has been found that $\sigma_{\rm pixel} \sim 1.7~\sigma_\gamma$, where
$\sigma_\gamma$ denotes the photon noise.
The value of  
$\sigma_{\rm pixel}$ and  $(\Delta F_{\rm pixel})_{\rm max}$ at the maximum
as well as their ratio are listed in Table~1.
The ratio $\sigma_{\rm pixel}/(\Delta F_{\rm pixel})_{\rm max}$ has mean value
$0.078 \pm
 0.026$. Therefore, for $Q=2$, we obtain $\delta_{\rm th} \simeq 15
\%$ for the detection
 of anomalies near the maximum.

For 'optimal image subtraction', the effective rms fluctuation can be pushed closer to
the photon-noise limit (Alard \& Lupton~\cite{Alard98}), yielding
$\sigma_{\rm pixel} \sim 1.2~\sigma_\gamma$, so that the detection threshold
reduces to 
$\delta_{\rm th} \simeq 10 \%$.

\section{Detection probability for planetary signals}

For $\mu$L{} events towards M31, the lens can be located in
the Milky Way halo, the M31 halo, or the M31 bulge. It
is almost impossible to discriminate among these different possible
locations of the lens from a single observed light curve, though
for a very small subset
of microlensing events it is possible to tell something
about the lens location (Han \& Gould~\cite{HG96}). 
Since we expect only those events for which the lens is in the
M31 bulge as being due to stars, 
we will consider only those events as potential targets for
a search for planetary anomalies.

As pointed out before, one also needs a
small impact parameter in order to produce an observable signal.
Therefore, we restrict our attention to events
that satisfy the following two conditions
\begin{enumerate}
\item $u_{0} < u_{\rm th} \equiv 0.1$; \footnote{For smaller $u_{\rm th}$, the detection probability will be
larger.}
\item{lens in the bulge or in the disk of the target galaxy.}
\end{enumerate}

Since we need more than one observed data point to be confident that we observe
a $\mu$L{} anomaly, we require an observable anomaly to 
deviate by more than 
$\delta_{\rm th}$ {\em and} during more than $t_{\rm E} / 100$,  
i.e.\ $\sim 7$~hours for a month-long event, therefore requiring some dense
sampling over the peak of the $\mu$L{} event.
The probability to detect a signal depends on the
projected separation $d$ between the star and the jupiter-like planet,
as defined in Sect.~\ref{miclens}.
Our calculation of the detection probability is similar
to the one done by Griest \& Safizadeh (\cite{GS98}), but
we use different detection criteria here.
For calculating the magnifications, we have used the approach developped by
Dominik (\cite{Dominik95}), released as 'Lens Computing Package (LCP)'.

\begin{figure}
\epsfig{figure=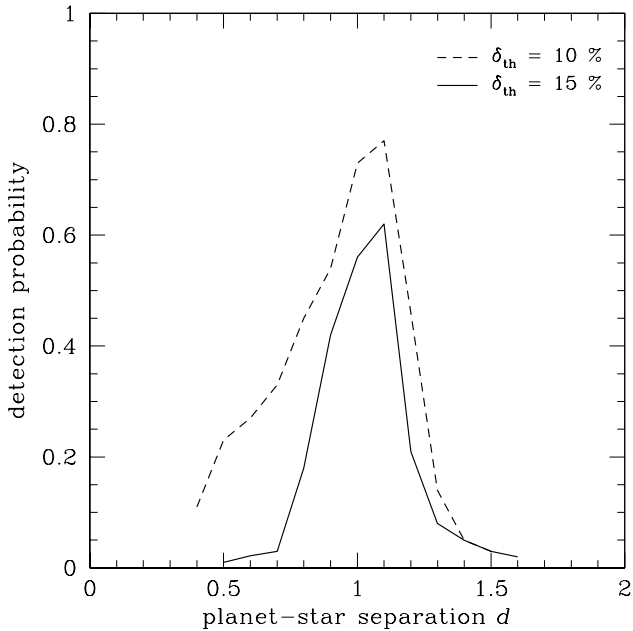,width=88mm}
\caption{The probability to see a deviation larger than $\delta_{\rm th} = 10\%$ or
$\delta_{\rm th} = 15\%$ caused by a
Jupiter-like planet ($q = 10^{-3}$) that lasts more than $t_{\rm E}/100 \sim 7~\mbox{hours}$ as a
function of the projected separation $d$ between star and planet in units of Einstein radii.}
\label{fig1}
\end{figure}

The ``cross
section'' of the central caustic depends strongly on the direction
of the source. Due to the elongated shape along the lens
axis, it has a maximum for trajectories orthogonal to this axis, and
a minimum for parallel trajectories. 
We have calculated the
largest impact parameter $u_{\rm max} \leq u_{\rm th}$ that satisfies
our detection criterium for several different source directions.
The detection probability for a planet for each of the
considered directions $\alpha$ is then simply given by
$P(\alpha) = u_{\rm max}(\alpha)/u_{\rm th}$, 
using the fact that the distribution of impact parameters
is approximately uniform for small impact parameters
for events from 
microlensing experiments towards unresolved star fields.
The final detection probability
has been calculated by averaging over the different trajectories.
The results are shown in Fig.~\ref{fig1}.

For both values of $\delta_{\rm th}$, 
there is some reasonable probability to detect planetary signals
for
planets in the lensing zone (i.e. the range of planetary position for
which the planetary caustics is within the Einstein ring of the major
component of the system, $0.618 \leq d \leq 1.618$). 
In agreement with previous work (Griest \& Safizadeh~\cite{GS98};
Dominik~\cite{Dominik99}), the detection probability reaches a maximum 
for planets located close to the Einstein ring of their parent star
(the caustic size increases towards $d \simeq 1$).
Averaged over the lensing zone, the detection probability is $\sim 20\%$ for
$\delta_{\rm th} = 15\%$ and $\sim 35\%$ for $\delta_{\rm th} = 10\%$. 
With a 2m-telescope, one can detected $\sim 400$ events
per year
towards the M31
bulge (Han~\cite{Han96}). 
Present-day microlensing surveys towards M31 are
still far away from such a theoretical limit, but the technique has
demonstrated to be successful, and fruitful developments can be
expected in the near future.
With $\sim 50\%$ of these events being due to M31 bulge lenses (Han~\cite{Han96}) and $\sim 50\%$ of these bulge lens events having $u_0 < 0.1$ 
(Baillon et al.~\cite{Baillon}), 
one
can expect to detect up to 35 anomalies caused by Jupiter-like planets
per year if every M31 bulge star has such a planet in its lensing zone.

To be able to observe and characterize the planetary anomaly, frequent 
observations (every few hours)
during the anomaly are necessary. Future observing programs 
towards M31 or other neigboring galaxies should take this into account.

\section{Extraction of planetary parameters}

There is a crucial difference between the detection of a signal that
is consistent with a planet and the detection of a planet, i.e.\ 
the determination of parameters that unambiguously characterize its nature.
In fact, it has been shown that 
the first microlensing event MACHO LMC-1 is
consistent with a planet (Rhie \& Bennett~\cite{RhieLMC1};
Alcock et al.~\cite{MACHObinary}). However, it appears to be consistent
with a binary lens of practically any mass ratio $q$ (Dominik \& Hirshfeld~\cite{DoHi2}), so that the existence of a planet cannot be claimed from this
event.

However, most of the papers about the detection of planets only show the possibility that
a signal that arises from a planet can be detected (Mao \& Paczy{\'n}ski~\cite{MP};
Griest \& Safizadeh~\cite{GS98}; Safizadeh et al.~\cite{Safplan}), while the question
about the
extraction of parameters has only been addressed by a few people. Dominik (\cite{Do:ND})
has stressed that this is complicated by several points: there may be several different
models that are consistent with the data, the fit parameters have finite uncertainties (in
particular blending strongly influences $t_{\rm E}$), and the physical lens parameters only
result on a stochastical basis using assumptions about galaxy dynamics.
Gaudi \& Gould (\cite{Gaudi97}) have shown that one needs frequent and precise observations to 
determine the mass ratio $q$ and the separation $d$ from type II anomalies.

However, it is more difficult to contrain these parameters in type I anomalies. Additional
complication arise because one does not obtain information about the
time separation between the main peak and the planetary peak,
there is a degeneracy between
 $d$ and $q$ (Dominik~\cite{Dominik99}), and
observed anomaly results from
 the combined action of all planets around the
lens star (Gaudi et al.~\cite{Naber98}).
 
Despite of the question whether $d$ and $q$ are well-determined, those parameters do not
give the mass of the planet $m$, nor its true separation $a$. 
Moreover,
an additional uncertainty enters because $d = a_{\rm p}/r_{\rm E}$ corresponds only to the
projected instantaneous separation $a_{\rm p}$. Using models for the galactic dynamics,
rather broad probability distributions for $a$ and $m$ result. 

However, as we stated before, photometric microlensing is the only method able to detect
signals of planets around stars in M31, so that if there is a way to find planets, this is the only
one. As we have shown, the prospects for detecting planetary signals are good. This means
that even if planets can be truly characterized in only a fraction of the events where signals
consistent with a planet can be detected, there is still a chance for being able to claim a
planet. Such a subset of events could e.g.\ consist of events where the source trajectory
crosses the caustic. Such caustic crossing events are likely to provide additional information.

A complete discussion of the extraction of planetary parameters is beyond
the scope of this paper and will be presented elsewhere
(Dominik \& Covone, in preparation).

\section{Summary and conclusions}

While microlensing is already the only method to detect planets around stars that are
at several kpc distance, namely by precise and frequent monitoring of
$\mu$L{} events towards
the Galactic bulge, future $\mu$L{} experiments towards nearby galaxies as M31 can
even push this distance limit much further.

Pixel lensing and difference image photometry have
demonstrated to be successful methods to search for
$\mu$L{} events towards unresolved star fields, and improvements 
are expected from the 'Optimal Image Subtraction (OIS)' technique
(Alard \& Lupton~\cite{Alard98}).

While AGAPE recently
reported the observation of the possible first anomalous $\mu$L{} event
towards M31 (Ansari et al.~\cite{Z1}), we
have shown that even planetary  
systems can give rise to measurable anomalies.
These planetary anomalies are due to passages of the source close to the central
caustic near the parent star, i.e.\ the detection channel discussed
by Griest \& Safizadeh (\cite{GS98}).

Using the estimate of Han (\cite{Han96}) that about 400 events per year
towards M31 can be detected with a 2m-telescope, we estimate that up to 35 jupiter-mass planets per year
can be detected if they exist frequently in the lensing zone around their parent star.
Following theoretical work by Gaudi \& Sackett (\cite{GS}),
PLANET (Albrow et al.~\cite{PLANET:O14}) and MPS and MOA (Rhie et al.~\cite{MPSMOA}) 
have recently published first results concerning the determination of the abundance
of planets from the absence of observed signals. From our estimates it follows that
future $\mu$L{} experiments towards
M31 can have the power to
yield strong constraints on the abundance of jupiter-mass planets.

\section*{Acknowledgements}

We gratefully thank V. Cardone, J. Kaplan, Y. Giraud-Heraud, P.
Jetzer and E. Piedipalumbo for stimulating discussions.
We also thank the referee for some remarks that 
helped in improving the paper.
RdR and AAM are finantially sustained by the M.U.R.S.T. grant PRIN97
``SIN.TE.SI''. 
GC has received support from the European Social Funds. 
The work of MD has been financed by a Marie Curie Fellowship (ERBFMBICT972457)
from the European Union.


\begin{thebibliography}{}

\bibitem[1998]{Alard98}
Alard C., Lupton R., 1998, ApJ 503, 325

\bibitem[1998]{PL}
Albrow M., et al. (The PLANET collaboration), 1998, ApJ 509, 687

\bibitem[1999]{PLANET:O14}
Albrow M., et al. (The PLANET collaboration), 1999, ApJ submitted, astro-ph/9909325

\bibitem[1996]{MACHO1}
Alcock C., et al. (The MACHO collaboration), 1996, ApJ 463, L67

\bibitem[1996]{MACHO2}
Alcock C., et al. (The MACHO collaboration), 1997, ApJ 479, 119

\bibitem[1999]{MACHObinary}
Alcock C., et al. (The MACHO and GMAN collaborations), 1999, ApJ submitted, astro-ph/9907369

\bibitem[1997]{Ansari97}
Ansari R., et al., 1997, A\&A 324, 843

\bibitem[1999]{Z1}
Ansari R., et al., 1999, A\&A 344, L49

\bibitem[1993]{Baillon}
Baillon P., Bouquet A., Giraud-H{\'e}raud Y., Kaplan J., 1993, A\&A 277, 1

\bibitem[1996]{Bennett96} 
Bennett D.P., Rhie S.H., 1996, ApJ 472, 660

\bibitem[1999]{SLOTT}
Bozza V., et al., 1999, SLOTT-AGAPE project. In: Proc. XLIII Congresso della
Societ{\`a}
 Astronomica Italiana,
 preprint astro-ph/9907162

\bibitem[1992]{Columbia}
Crotts A.P.S., 1992, ApJ 399, L43

\bibitem[1996]{CT}
Crotts A.P.S., Tomaney A.B., 1996, ApJ 473, L87

\bibitem[1999]{MEGA}
Crotts A.P.S., Uglesich R., Gyuk G., 1999, MEGA, a Wide-Field Survey of Microlensing in M31. 
In: Brainerd T., Kochanek C.S. (eds.) Gravitational Lensing:
Recent Progress and Future Goals. ASP Conf. Ser., ASP, San Francisco, in press,
preprint astro-ph/9910552

\bibitem[1995]{Dominik95}  
Dominik M., 1995, A\&AS 109, 597

\bibitem[1997]{Do:ND}
Dominik M., 1997, The extraction of information from binary and planetary
lensing light curves. In: Proc. of the 3rd International Workshop on Gravitational
Microlensing Surveys, Notre Dame, Indiana, USA

\bibitem[1998]{Dominik98}  
Dominik M., 1998, A\&A 330, 963

\bibitem[1999]{Dominik99}  
Dominik M., 1999, A\&A 349, 108

\bibitem[1996]{DoHi2}  
Dominik M., Hirshfeld A.C., 1996, A\&A 313, 841

\bibitem[1998]{Sahu}  
Dominik M., Sahu K.C., 1998, ApJ submitted, astro-ph/9805360

\bibitem[1999]{PL2}
Dominik M., et al. (The PLANET collaboration), 1999, Physics and Chemistry of the Earth, submitted, 
astro-ph/9910465

\bibitem[1993]{Erdl}
Erdl H., Schneider P., 1993, A\&A 268, 453

\bibitem[1997]{Gaudi97}  
Gaudi B.S., Gould A., 1997, ApJ 486, 85

\bibitem[1998]{Naber98} 
Gaudi B.S., Naber R.M., Sackett P.D., 1998, ApJ 502, L33

\bibitem[2000]{GS}
Gaudi B.S., Sackett P.D., ApJ, in press, astro-ph/9904339

\bibitem[1992]{Gould92}  
Gould A., Loeb A., 1992, ApJ 396, 104

\bibitem[1996]{Gould96}  
Gould A., 1996, ApJ 470, 201

\bibitem[1998]{GS98}  
Griest K., \& Safizadeh N., 1998, ApJ 500, 37

\bibitem[1996]{Han96}  
Han C., 1996, ApJ 472, 108

\bibitem[1996]{HG96}  
Han C., Gould A., 1996, ApJ 473, 230

\bibitem[1999]{MOA}
Hearnshaw J.B., et al., 1999, Photometry of pulsating stars in the Magellanic Clouds as
observed in the MOA project. In: Szabados L., Kurtz D. (eds.) The Impact of large-scale Surveys
on Pulsating Star Research -- IAU Colloquium 176, ASP Conf. Ser, ASP, San Francisco

\bibitem[1698]{Huyghens}  
Huygens C., 1698, The Celestial Worlds Discovered

\bibitem[1998]{Jetzer98}  
Jetzer P., 1998, Gravitational Microlensing. In:
Marino A.A., et al. (eds.) Topics in Gravitational Lensing,
Bibliopolis, Napoli

\bibitem[1998]{Kaplan97}  
Kaplan J., 1998, Pixel Lensing. In:
Marino A.A., et al. (eds.) Topics in Gravitational Lensing,
Bibliopolis, Napoli

\bibitem[1991]{MP}  
Mao S., Paczy\'nski B., 1991, ApJ 374, L37

\bibitem[1996]{planet2}
Marcy G.W., Butler R.P., 1996, ApJ 464, L147

\bibitem[1998]{Marcy98} 
Marcy G.W., Butler R.P., 1998, ARA\&A 36, 57

\bibitem[1995]{planet1}
Mayor M., Queloz D., 1995, Nat 378, 355

\bibitem[1998]{Mel} 
Melchior A.-L., 1998, AGAPE -- Summary and Prospects. In:
Spooner N. (eds.) Proc. 2nd International 
Workshop on the Identification of Dark Matter, Buxton, England

\bibitem[1986]{Pac86}  
Paczy\'nski B., 1986, ApJ 304, 1

\bibitem[1996]{Pac-rev}  
Paczy\'nski B., 1996, ARA\&A 34, 419

\bibitem[1998]{EROS}
Palanque-Delabrouille N., et al. (The EROS collaboration), 1998, A\&A 332, 1

\bibitem[1996]{RhieLMC1}
Rhie S.H., Bennett D.P., 1996, Nucl. Phys. Proc. Suppl. 51, 86

\bibitem[1999a]{MPS}
Rhie S.H., et al. (The Microlensing Planet Search collaboration), 1999, ApJ 522, 1037

\bibitem[1999b]{MPSMOA}
Rhie S.H., et al. (The MPS and MOA collaborations), preprint astro-ph/9905151

\bibitem[1997]{Roulet97}  
Roulet E., Mollerach S., 1997, Phys. Rep. 279, 67

\bibitem[1999]{Safplan}  
Safizadeh N., Dalal N., Griest K., 1999, ApJ 522, 512

\bibitem[1994]{SahuNat}
Sahu K.C., 1994, Nat 370, 275

\bibitem[1986]{SchneiWei}
Schneider P., Wei{\ss} A., 1986, A\&A 164, 237

\bibitem[1999]{planetencyc}
Schneider J., 1999, {\tt http://www.obspm.fr/planets}

\bibitem[1996]{TC}
Tomaney A.B, Crotts A.P.S., 1996, AJ 112, 2872

\bibitem[1997]{OGLE}
Udalski A., Kubiak M., Szyma{\'n}ski M., Acta Astron. 47, 319

\bibitem[1992]{Wols1}  
Wolszczan A., Frail D.A., 1992, Nat 355, 145 

\bibitem[1994]{Wols2}  
Wolszczan A., 1994, Sci 264, 538 

\end{thebibliography}
\end{document}